\documentclass{emulateapj}

\begin{document}

\title{Metallicity Effects on Mid-Infrared Colors and the 8\micron\ PAH
Emission in Galaxies}

\shorttitle{Metallicity Effects on MIR Galaxy Colors}

\author{C. W. Engelbracht\altaffilmark{1}, K. D. Gordon\altaffilmark{1},
G. H. Rieke\altaffilmark{1}, M. W. Werner\altaffilmark{2}, D. A.
Dale\altaffilmark{3}, and W. B. Latter\altaffilmark{4}}

\altaffiltext{1}{Steward Observatory, University of Arizona, Tucson, AZ 85721;
cengelbracht@as.arizona.edu}
\altaffiltext{2}{Jet Propulsion Laboratory, MC 264-767, 4800 Oak Grove Drive,
Pasadena, CA 91109}
\altaffiltext{3}{Department of Physics and Astronomy, University of Wyoming,
Laramie, WY 82071}
\altaffiltext{4}{Spitzer Science Center, California Institute of Technology,
Pasadena, CA 91125}

\begin{abstract}

We examine colors from 3.6\micron\ to 24\micron\ as a function of metallicity
(O/H) for a sample of 34 galaxies.  The galaxies range over 2 orders of
magnitude in metallicity. They display an abrupt shift in the 8\micron\ to
24\micron\ color between metallicities $1/3$ to $1/5$ of the solar value.  The
mean 8\micron\ to 24\micron\ flux density ratio below and above $12 + log
(O/H) = 8.2$ is $0.08 \pm 0.04$ and $0.70 \pm 0.53$, respectively.  We use
mid-infrared colors and spectroscopy to demonstrate that the shift is
primarily due to a decrease in the 8\micron\ flux density as opposed to an
increase in the 24\micron\ flux density.  This result is most simply
interpreted as due to a weakening at low metallicity of the mid-infrared
emission bands usually attributed to PAHs (polycyclic aromatic hydrocarbons)
relative to the small-grain dust emission.  However, existing empirical
spectral energy distribution models cannot account for the observed
short-wavelength (i.e., below 8\micron) colors of the low-metallicity galaxies
merely by reducing the strength of the PAH features; some other emission
source (e.g., hot dust) is required.

\end{abstract}


\section{Introduction}

Early galaxies must have formed at a very low metallicity.  Such galaxies
likely had different properties from typical galaxies observed in the local
universe, where generations of star formation have enriched the ISM
(interstellar medium).  These early galaxies are difficult to study because
they are so distant.  As a result, nearby low-metallicity galaxies have
generated considerable interest as local analogs of young galaxies at high
redshift \citep[see, for example, the review by][]{kun00}.  

Understanding the infrared properties of nearby low metallicity galaxies will
help guide interpretation of {\it Spitzer} observations of high-redshift ones
\citep[e.g.,][]{faz04,yan04,lag04}. An important aspect of infrared galaxy
spectral energy distributions (SEDs) is the behavior between 3 and 24\micron,
a range that includes emission from PAH (Polycyclic Aromatic Hydrocarbon)
molecules and warm dust.  At redshifts between 1 and 3, the PAH features move
through the MIPS 24um band. {\it Ad hoc} adjustments to this spectral region
in SED models \citep{lag04} have been required to fit the 24\micron\ number
counts \citep{pap04,mar04,cha04}. \citet{lia04} show that infrared luminous
galaxies at z $\sim$ 0.6 are only $\sim$ 50\% as metal-rich as their local
analogs. Therefore, interpreting the behavior of galaxies detected with {\it
Spitzer} to substantially higher redshifts (e.g., Per\'ez-Gonz\'alez et al.
2005, submitted to ApJ) depends critically on understanding the effects of
metallicity on the 24$\mu$m flux densities for 1 $<$ z $<$ 3.  

Low-metallicity galaxies are difficult to find:  I~Zw~18, the first extreme
example identified \citep{sea72} remains the lowest-metallicity galaxy yet
found, despite 3 decades of searching.  Low-metallicity galaxies have had
little or no star formation up to the present epoch and are therefore of very
low luminosity.  As a result, most existing compilations of the MIR properties
of galaxies are restricted to high-metallicity, high-luminosity targets
\citep[e.g.,][]{gen98,lu03}.  Where low-metallicity galaxy spectra have been
obtained, though, they differ dramatically from their higher-metallicity
cousins in their infrared properties, with hotter large-grain dust emission
and weak PAH emission \citep{sau90,cal00,gal03,con00,
thu99,hou04b,roc91,mad02}.  

This work greatly expands the sample of low-metallicity galaxies observed in
the infrared, to explore the effects of metallicity on the infrared properties
of galaxies in a systematic way.  The larger sample has allowed us to pinpoint
the metallicity at which the transition from weak to strong PAH emission
occurs.

\section{Observations and Data Reduction}
\label{sec:data}

The new data presented here were obtained using IRAC and MIPS.  The IRAC data
are all standard pipeline reductions using the most recent versions available
in the archive, ranging from 9.5.0 to 10.5.0.  The MIPS data were reduced
using the MIPS DAT \citep[Data Analysis Tool;][]{gor05}.

Photometry was performed using the ``imexam'' and ``imstat'' tasks in
IRAF\footnote{IRAF is distributed by the National Optical Astronomy
Observatories, which are operated by the Association of Universities for
Research in Astronomy, Inc., under cooperative agreement with the National
Science Foundation.}.  Most sources in the sample are compact; in general,
apertures large enough to encompass all the flux were used for MIPS, while the
standard aperture with a radius of 10 pixels (12\arcsec) was used for IRAC, as
suggested in the IRAC data handbook.  For the extended galaxies, matching
apertures that include the whole galaxy were used for MIPS and IRAC, and
extended-source corrections have been applied to the IRAC measurements as
described in the IRAC data handbook.  No color corrections were applied, which
should not have a strong impact on the results since the corrections listed in
the IRAC and MIPS handbooks are typically only a few percent.

To increase the number of metal-rich galaxies in the sample, we have
supplemented our measurements with results from the literature.  Where
IRAC and MIPS measurements were available, they have been taken directly from
the sources indicated in Table~\ref{tab1}.  We have synthesized IRAC and MIPS
measurements from an IRS \citep[Infrared Spectrograph;][]{hou04a} measurement
of NGC~7714 by \citet{bra04} by convolving the spectrum with the IRAC
8\micron\ and MIPS 24\micron\ spectral response curves available on the SSC
({\it Spitzer} Science Center) website.

Finally, we have added several measurements from MSX (Midcourse Space
Experiment) by \citet{kra02} and IRAS (Infrared Astronomical Satellite) by
\citet{ric88}.  The MSX Band A (8.28\micron) and IRAC 8\micron\ flux densities
measured for two galaxies (M~101, K.  Gordon, in preparation; M~83, C.
Engelbracht, in preparation) are very similar (within 20\% in both cases), so
we have made no correction to the MSX measurements used in this paper.  We
have made an empirical color correction from the IRAS 25\micron\ band to the
MIPS 24\micron\ band by comparing the measured flux densities of NGC~7331
\citep{reg04}, M~81 \citep{gor04}, NGC~55 \citep{eng04}, and M~101 (K.
Gordon, in preparation).  The measured 24\micron\ to 25\micron\ ratios ranged
from 0.81 to 0.94, so we applied the average value, 0.87, to all the IRAS
measurements incorporated into the flux density ratio measurements in
Table~\ref{tab1}.

The resulting sample is a heterogeneous collection of star-forming galaxies
without strong active nuclei.  The global photometry measurements make this
sample suitable for comparison to high-redshift samples
\citep[e.g.,][]{faz04}, where only flux measurements integrated over the whole
galaxy are feasible.  The metallicity quoted for each galaxy generally applies
to a much smaller beam than for the infrared data. The outer regions of each
galaxy should have lower metallicity than the nuclei.  Thus, the beam mismatch
would, if anything, weaken the trend of infrared properties we see clearly in
the suite of measurements. We conclude it is not an impediment to this study.

\section{Results}
\label{sec:results}

To explore the contribution of PAH emission to the 8\micron\ band, we compare
the 8\micron\ measurements to bands at both shorter and longer wavelengths.
The longer-wavelength 24\micron\ band is already dominated by emission by dust
and so we use it directly, but the shorter-wavelength 5.8 and 4.5\micron\
bands can contain significant contributions from PAH emission and starlight,
respectively.  To avoid contamination from PAH features, we do not use the
5.8\micron\ band to derive the dust continuum measurement.  We instead use the
4.5\micron\ band, from which we subtract the stellar component as follows: We
assume the 3.6\micron\ band is dominated by starlight \citep[which may result
in an underestimate of the 4.5\micron\ dust emission by a few percent;
cf.][]{hel04} and multiply the 3.6\micron\ measurement by the 4.5/3.6\micron\
ratio expected for the stellar population, derived from Starburst99
\citep{lei99} data files available on the
web\footnote{http://www.stsci.edu/science/starburst99/}.  The ratio predicted
by that model is typically 0.53 to 0.61, depending on star formation history
and metallicity.  We adopt the average value of 0.57 and assign to it an
uncertainty of 7\%.  Hereafter, we refer to this ratio as $\alpha$.  Our
adopted value of $\alpha$ is similar, but not identical to, the value of 0.47
in the template elliptical galaxy spectrum used by \citet{lu03} to measure the
hot dust component in a sample of galaxies.  This will have some effect on our
measurement of the 4.5\micron\ dust component for specific galaxies but does
not affect the conclusions we draw from a diagnostic plot, as we discuss
below.  The scaled 3.6\micron\ measurement is then subtracted from the
4.5\micron\ measurement.  For the galaxies in this sample, the scaled
3.6\micron\ measurement ranges from 26\% to 96\% of the 4.5\micron\
measurement, with an average value of 70\%.

We did not subtract the stellar contribution from the 8\micron\ and 24\micron\
measurements because the correction is small for these galaxies and has no
impact on our conclusions.  Using the same Starburst99 models as above, we
derive that the stellar contribution at 8\micron\ ranges from 1\% to 30\% with
an average of 10\% and the contribution at 24\micron\ is negligible, less than
1\% on average.

The 8\micron\ measurements compared to the short and long wavelength bands are
summarized in Table~\ref{tab1}, where the ratio to the 4.5\micron\ band is
referred to as $R_1$ and the ratio to the 24\micron\ band as $R_2$. Increasing
levels of PAH emission will result in decreasing values of $R_1$ and
increasing values of $R_2$.  We note that any correction for dust in the
3.6\micron\ band will tend to increase the calculated 4.5\micron\ dust
emission and therefore increase the value of $R_1$ for those galaxies without
strong PAH emission.

\begin{figure}
\plotone{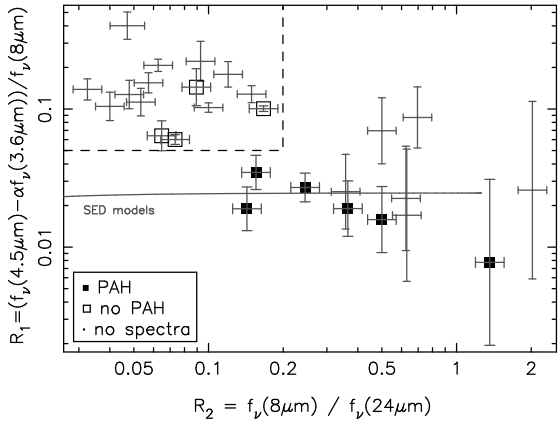}
\caption{MIR colors of galaxies with a range of metallicities.  The 8\micron\
band is compared to both shorter (4.5\micron, with stellar emission subtracted
as described in the text) and longer (24\micron) wavelengths.  Galaxies with
known PAH features are indicated by filled markers, while galaxies known to
lack PAH features are indicated by open markers.  Error bars include a 10\%
relative calibration uncertainty between the IRAC and MIPS instruments and a
7\% uncertainty on $\alpha$, in addition to photometric uncertainties.  The
dashed line indicates our chosen separation between PAH and non-PAH galaxies
in this color space.  The range of points covered by the \citet{dal01a} SED
models is shown as a solid line.}
\label{fig1}
\end{figure}

The color measurements are presented graphically in Figure~\ref{fig1}.  The
figure shows that galaxies with high values of $R_1$ tend to have low values
of $R_2$, while those galaxies with a value of $R_1$ in the range of 0.01 to
0.09 \citep[similar to our galaxy; cf.][]{lu04} tend to have larger values of
$R_2$, although with a large dispersion.  We find that the separation of
galaxies in this plot persists if we change our assumptions about how to
compute the 4.5\micron\ dust continuum, e.g., by changing the value of
$\alpha$ to 0.47 to match the value used by \citet{lu03} or by using
near-infrared data (such as the J band, which should not have a contribution
from hot dust) to determine the stellar contribution to the 4.5\micron\ band.
This trend is consistent with the spectroscopic results summarized in the
``PAH'' column of Table~\ref{tab1}, which demonstrate that all the galaxies
with spectra free of PAH emission cluster in the upper-left portion of the
diagram.  

To interpret these results, we used the SED models of \citet{dal01a}.  To
compute photometry from the models (which do not contain a stellar emission
component), we convolved them with the same spectral
response curves discussed in \S~\ref{sec:data}.  The various flux density
ratios were then computed as for the data.  The emission in the IRAC bands
derived from these models is dominated by a fixed PAH spectral template, and
as a result the models have a nearly constant value of $R_1$, around 0.025.
The models cover a range of $R_2$ values that depends on the temperature of
the thermally pulsing grains, along the locus of points {\it outside} the
region to the upper left occupied by the PAH-free galaxies.  Thus, the
weakness of the 8\micron\ emission relative to the other bands appears to be
due to the weakness of the PAH emission relative to the emission from small
dust grains.

\begin{figure}
\plotone{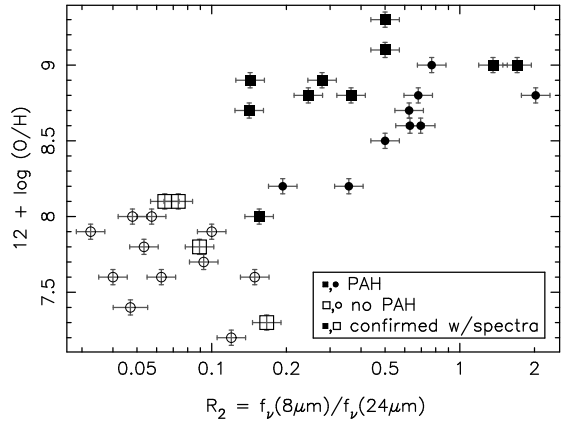}
\caption{Galaxy metallicity as a function of the 8 to 24\micron\ color.
Galaxies with colors or spectra that indicate that they have 8\micron\ PAH
features are displayed as filled markers, while galaxies which lack the
8\micron\ PAH feature are shown as open markers.  Squares and circles denote
measurements with and without spectroscopic confirmation, respectively.  The
error bars on the 8/24\micron\ ratio are the same as in Figure~\ref{fig1},
while the error bars on the metallicity are typically 0.05 or less
\citep[cf.][]{kob96} and were all assigned an uncertainty of 0.05.  Note that
this figure does not require IRAC data (some of the 8\micron\ measurements
were made by MSX) and thus contains more points than Figure~\ref{fig1}.}
\label{fig2}
\end{figure}

Figure~\ref{fig2} shows the 8-to-24$\mu$m color vs. metallicity \citep[the
solar metallicity on this scale is 8.7;][]{all01}.  We label the galaxies
according to the lack or presence of 8\micron\ PAH emission using the color
separation from Figure~\ref{fig1}. They separate very cleanly according to
metallicity, with a narrow transition region around $1/3$ to $1/5$ solar
metallicity.  One additional galaxy, NGC 1569, has been shown to have weak PAH
emission \citep{lu03} and has an O/H-based metallicity 25\% of solar,
consistent with this overall behavior.

The boundary between metallicities with and without PAH emission is
surprisingly sharp, and it seems likely that future studies will show more
scatter. For example, there can be significant variations in metallicity
within these systems \citep[e.g., a factor of three is indicated for
SBS~0335-052;][]{hou04b}.  Our characterization of this parameter with a
single value is an oversimplification. In addition, "metallicity" includes
enrichment of the ISM with a broad variety of elements, produced in different
ways. Oxygen is produced predominantly in very massive stars; use of other
elements produced in other ways as metallicity indicators may yield different
behavior. Nonetheless, there is a robust trend toward low PAH emission in
galaxies with relatively unpolluted ISMs. 

Why do the PAH features become weaker in low-metallicity galaxies?  We reject
extinction as a possible cause, both because the effect would have to
(counterintuitively) become stronger in galaxies with lower heavy-element
abundance (and presumably less dust) and because the direct measurement of
extinction in the low-metallicity galaxy SBS~0335-052 shows it to be optically
thin in the MIR \citep{hou04b}.  It is also unlikely that starburst activity
(i.e., the intensity of the star formation) is a major contributor to the
effect:  the correlation coefficient between $R_2$ and the 24\micron\ surface
brightness (an indicator of the specific star formation rate, which we
estimated from the integrated flux densities and the visible extent of the
galaxies in the sample) is much lower than the correlation coefficient between
$R_2$ and the metallicity (0.26 vs.  0.79).  Also, while the
well-known starburst galaxies in our sample (e.g., NGC~253, NGC~7714), do tend
to have lower $R_2$ values than other galaxies at similar metallicity, none of
them have as small an $R_2$ value as the lowest metallicity galaxies.  One possible reason is the
destruction of the PAH carrier in the harsh radiation field of a
low-metallicity galaxy \citep{pla02}.  Another possibility is that these
galaxies are truly young and the PAH molecules have simply not had time to
form \citep{dal01b}.  For example, it is well known that the ISM of a galaxy
must be significantly polluted with oxygen (produced copiously in very massive
stars) before significant amounts of carbon are produced (by stars of a few
M$_\odot$) --- this effect may delay PAH production (E. Dwek, private
communication).  The low-metallicity galaxies may simply lack the carbon-rich
AGB stars required to form the PAH molecules \citep{lat91}. 

In addition, we find that the PAH features included in the \citet{dal01a}
models fail to account for the high values of $R_1$ prevalent at low
metallicity (i.e., below 8.0).  The emission by dust in the models below
10\micron\ is several orders of magnitude lower than the emission from the PAH
template (which does not vary), so the simple remedy to the models of
weakening the PAH spectrum does not allow them to match the low-metallicity
data.  The value of $R_1$ derived from the model depends only on the shape of
the PAH template spectrum and does not match the low-metallicity galaxies
shown in Figure~\ref{fig1} (i.e., the model $R_1$ value plotted in
Figure~\ref{fig1} does not shift as the PAH features are weakened).  Some
other source of emission must be dominating in these sources, such as hot dust
in the 4.5\micron\ band \citep[cf.][]{lu03}.

\section{Conclusions}

We have combined new MIR measurements from the {\it Spitzer Space Telescope}
with results from the literature to show that the 8\micron\ to 24\micron\ flux
density ratio in star-forming galaxies depends strongly on metallicity.  The
8\micron\ to 24\micron\ color changes markedly between $1/3$ to $1/5$ solar
metallicity --- the mean ratio below $1/3$ solar metallicity is $0.08 \pm
0.04$, while the ratio at higher metallicity is $0.70 \pm 0.53$.  A MIR
color-color diagram which compares the 8\micron\ band to dust continuum
measurements at 4.5 and 24\micron\ shows that the change in the ratio is
predominantly due to a decrease in the 8\micron\ emission.  We interpret this
result as a weakening of the PAH features at low metallicity, an
interpretation consistent with the spectroscopic evidence available in the
literature.  

We show that existing empirical SED models cannot account for the color change
merely by weakening the PAH component of the models.  In addition to the
dependence of PAH strength on metallicity, SED models must account for another
source of emission (e.g., hot dust) to explain the IRAC observations of
low-metallicity galaxy colors.

The shift in 8 to 24\micron\ ratio occurs at a fairly high metallicity and
will thus affect the spectra of many galaxies.  Future SED models 
to interpret {\it Spitzer} measurements at high redshift must take
the dependence of metallicity on redshift into account.

\acknowledgments

We thank Eli Dwek for helpful discussions, and Nanyao Lu and Eiichi Egami for
comments which improved this paper.  This work is based in part on
observations made with the {\it Spitzer Space Telescope}, which is operated by
the Jet Propulsion Laboratory, California Institute of Technology under NASA
contract 1407. Support for this work was provided by NASA through Contract
Number 960785 issued by JPL/Caltech.

\clearpage

\begin{deluxetable}{lllllccccc}
\tablewidth{0pt}
\tablecaption{Galaxy Colors and Metallicities\label{tab1}}
\tablehead{\colhead{Galaxy} &
\colhead{f$_\nu$(4.5\micron)} & \colhead{f$_\nu$(24\micron)} &
\colhead{$R_1$\tablenotemark{a}} &
\colhead{$R_2$\tablenotemark{b}} & \colhead{Z\tablenotemark{c}} &
\colhead{PAH}\tablenotemark{d} & \multicolumn{3}{c}{References}
\\ \colhead{} & \colhead{Jy} & \colhead{Jy} & \colhead{} & \colhead{} &
\colhead{} & \colhead{} &
\colhead{flux densities\tablenotemark{e}} & \colhead{Z} & \colhead{PAH}}
\startdata
I Zw 18 & 3.4e-4 & 5.5e-3 & 0.18 & 0.12 & 7.2 & ... & 1 & 2 & ... \\
SBS 0335-052 & 1.5e-3 & 6.6e-2 & 0.10 & 0.17 & 7.3 & no & 1 & 3 & 4 \\
HS 0822+3542 & 1.45e-4 & 2.7e-3 & 0.40 & 0.06 & 7.4 & ... & 1 & 5 & ... \\
Tol 1214-277 & 7.2e-5 & 5.5e-3 & 0.10 & 0.04 & 7.6 & ... & 1 & 2 & ... \\
Tol 65 & 3.6e-4 & 1.5e-2 & 0.21 & 0.06 & 7.6 & ... & 1 & 2 & ... \\
Tol 2138-405 & 2.4e-3 & 5.7e-2 & 0.13 & 0.15 & 7.6 & ... & 1 & 6 & ... \\
VII Zw 403 & 2.4e-3 & 2.8e-2 & 0.22 & 0.09 & 7.7 & ... & 1 & 7 & ... \\
Mrk 153 & 1.45e-3 & 2.9e-2 & 0.14 & 0.08 & 7.8 & no & 1 & 8 & 9 \\
UM 461 & 5.5e-4 & 3.0e-2 & 0.11 & 0.05 & 7.8 & ... & 1 & 2 & ... \\
Haro 11 & 3.2e-2 & 1.9e+0 & 0.10 & 0.10 & 7.9 & ... & 1 & 10 & ... \\
NGC 4861 & 3.2e-3 & 2.8e-1 & 0.14 & 0.03 & 7.9 & ... & 1 & 11 & ... \\
Mrk 1450 & 9.2e-4 & 4.8e-2 & 0.13 & 0.05 & 8.0 & ... & 1 & 12 & ... \\
UM 448 & 1.1e-2 & 5.6e-1 & 0.03 & 0.16 & 8.0 & yes & 1 & 12 & 9 \\
UM 462 & 2.4e-3 & 1.1e-1 & 0.15 & 0.06 & 8.0 & ... & 1 & 2 & ... \\
Mrk 930 & 2.3e-3 & 1.7e-1 & 0.06 & 0.06 & 8.1 & no & 1 & 12 & 9 \\
II Zw 40 & 1.1e-2 & 1.5e+0 & 0.06 & 0.07 & 8.1 & no & 1 & 2 & 13 \\
NGC 4670 & 1.5e-2 & 2.1e-1 & 0.03 & 0.36 & 8.2 & ... & 1 & 14 & ... \\
NGC 1156 & ... & 4.5e-1 & ... & 0.19 & 8.2 & ... & 15 & 16 & ... \\
NGC 0598 & ... & 5.0e+1 & ... & 0.50 & 8.5 & ... & 17,18 & 19 & ... \\
NGC 3077 & 2.76e-1 & 1.5e+0 & 0.02 & 0.48 & 8.6 & ... & 1 & 20 & ... \\
NGC 300 & 8.8e-1 & 2.3e+0 & 0.09 & 0.70 & 8.6 & ... & 21 & 19 & ... \\
NGC 2537 & 4.1e-2 & 2.4e-1 & 0.02 & 0.62 & 8.7 & ... & 1 & 22 & ... \\
NGC 7714 & ... & 2.4e+0 & ... & 0.14 & 8.7 & yes & 23 & 20 & 23 \\
NGC 2782 & 3.8e-2 & 9.6e-1 & 0.02 & 0.36 & 8.8 & yes & 1 & 14 & 24 \\
NGC 4194 & 6.5e-2 & 3.1e+0 & 0.03 & 0.25 & 8.8 & yes & 1 & 20 & 25 \\
NGC 5457 & ... & 1.0e+1 & ... & 0.68 & 8.8 & ... & 26 & 19 & ... \\
NGC 3031 & 6.5e+0 & 4.4e+0 & 0.03 & 2.02 & 8.8 & ... & 27,28 & 19 & 29 \\
He 2-10 & 6.0e-2 & 4.9e+0 & 0.02 & 0.14 & 8.9 & yes & 1 & 11 & 24 \\
NGC 253 & ... & 1.4e+2 & ... & 0.28 & 8.9 & yes & 17,30 & 19 & 31 \\
NGC 5055 & ... & 6.1e+0 & ... & 0.77 & 9.0 & ... & 17,30 & 19 & ... \\
NGC 7331 & 9.5e-1 & 3.6e+0 & 0.01 & 1.36 & 9.0 & yes & 32 & 19 & 33 \\
NGC 224 & ... & 9.4e+1 & ... & 1.70 & 9.0 & yes & 17,30 & 19 & 34 \\
NGC 5236 & ... & 4.2e+1 & ... & 0.50 & 9.1 & yes & 1,30 & 19 & 29 \\
NGC 2903 & 1.2e-1 & 2.2e+0 & 0.02 & 0.50 & 9.3 & yes & 1 & 11 & 9 \\
\enddata
\tablenotetext{a}{$R_1 = [f_\nu(4.5\micron) - \alpha f_\nu(3.6\micron)] /
f_\nu(8\micron)$, where $\alpha$ is as described in the text.}
\tablenotetext{b}{$R_2 = f_\nu(8\micron) / f_\nu(24\micron)$}
\tablenotetext{c}{$12 + log (O/H)$}
\tablenotetext{d}{The PAH designation is based on results reported in the cited works or on visual
inspection of the spectra.}
\tablenotetext{e}{A single entry refers to both IRAC and MIPS
measurements together.  Two entries indicate separate sources for the
measurements, with the IRAC reference followed by the MIPS
reference.}
\tablerefs{(1) This paper; (2) \citet{kob96}; (3) \citet{izo99}; (4)
\citet{hou04b}; (5) \citet{kni00}; (6) \citet{mas94}; (7) \citet{izo97}; (8)
\citet{kun85}; (9) Engelbracht et~al., in preparation; (10) \citet{ber02};
(11) \citet{kob99}; (12) \citet{gus00}; (13) \citet{mad02}; (14)
\citet{hec98}; (15) \citet{pah04}; (16) \citet{lis98}; (17) MSX Band A
measurements by \citet{kra02}; (18) \citet{hin04}; (19) \citet{gar02}; (20)
\citet{sto94}; (21) \citet{hel04}; (22) J. Moustakes, private communication;
(23) IRS spectroscopy by \citet{bra04} convolved with IRAC and MIPS response
curves as described in the text; (24) \citet{roc91}; (25) \citet{lu03}; (26)
K.  Gordon, in preparation; (27) \citet{wil04}; (28) \citet{gor04}; (29)
\citet{rig99}; (30) IRAS measurements by \citet{ric88} transformed to MIPS
24\micron\ measurements as described in the text; (31) \citet{dev04}; (32)
\citet{reg04}; (33) \citet{smi04}; (34) \citet{ces98}.}
\end{deluxetable}

\end{document}